\documentclass[final]{anthology-ch} 

\usepackage{booktabs}
\usepackage{graphicx}

\usepackage{amsmath}
\usepackage{multirow}
\PassOptionsToPackage{hyphens}{url}
\usepackage{hyperref}

\usepackage{xcolor}      
\usepackage{tcolorbox}
\tcbuselibrary{skins}    
\usepackage{etoolbox}    

\definecolor{myred}{RGB}{173, 0, 88}

\newcommand{\MyBox}[6]{%
\begin{tcolorbox}[
    colback=myred!3!white,
    colframe=myred,
    title=#2,
    fonttitle=\bfseries,
    boxrule=0.8pt,
    arc=2pt,
    left=6pt,
    right=6pt,
    top=6pt,
    bottom=6pt,
    enhanced
]
\begin{tcolorbox}[
    colback=white,
    colframe=myred!60,
    boxrule=0.4pt,
    arc=1pt,
    left=4pt,
    right=4pt,
    top=4pt,
    bottom=4pt,
    sidebyside,
    sidebyside gap=4mm,
]
#3
\tcblower
#4
\end{tcolorbox}
\vspace{4pt}
\ifstrempty{#6}
{
    \begin{tcolorbox}[
        colback=white,
        colframe=myred!60,
        boxrule=0.4pt,
        arc=1pt,
        left=4pt,
        right=4pt,
        top=4pt,
        bottom=4pt,
    ]
    #5
    \end{tcolorbox}
}
{
    \begin{tcolorbox}[
        colback=white,
        colframe=myred!60,
        boxrule=0.4pt,
        arc=1pt,
        left=4pt,
        right=4pt,
        top=4pt,
        bottom=4pt,
        sidebyside,
        sidebyside gap=4mm,
    ]
    #5
    \tcblower
    #6
    \end{tcolorbox}
}
\end{tcolorbox}
}

\title{Retrieving Patristic References and Modelling Topics in Bullinger’s Reformation Correspondence}

\author[1]{Lena Pickartz}[
  orcid=0009-0000-8185-3002
]

\author[1]{Philipp Schaer}[
  orcid=0000-0002-8817-4632
]

\author[2]{Phillip B.~Ströbel}[
  orcid=0000-0003-2063-5495
]

\affiliation{1}{TH Köln -- University of Applied Sciences, Cologne, Germany}
\affiliation{2}{Department of History, University Zurich, Switzerland}

\keywords[eng]{intertextuality detection, hybrid retrieval, topic modelling, historical NLP}

\pubyear{2026}
\pubvolume{1}
\pagestart{1}
\pageend{1}
\paperorder{1}
\conferencename{Proceedings of Conference XXX}
\conferenceeditors{Editor1 Editor2}
\doi{00000/00000}  
\customhead{}  

\begin{document}

\maketitle

\begin{abstract}
The correspondence of Heinrich Bullinger (1504--1575), Swiss Reformer and Antistes of Zurich, documents the political and religious debates of the Reformation, including extensive appeals to the Church Fathers as authoritative sources. Detecting such references computationally remains largely unexplored. We investigate dense and sparse retrieval for patristic reference detection, and metadata-informed topic modelling as a proxy for the topics and ideas discussed. Prior work on Latin intertextuality has fine-tuned dense retrievers for their respective target domain, and we test how well these models transfer to Bullinger's correspondence. We compare lexical (BM25), dense (bi-encoders), and hybrid methods on explicit and implicit references. A Latin encoder fine-tuned on biblical reuse performs best among the bi-encoders (82\% $Recall@20$ on explicit and 16\% on implicit references), while BM25 outperforms dense approaches on explicit citations (94\%, $R@20$). Because BM25 and dense retrieval recover disjoint sets of implicit references, fusing their rankings raises implicit $R@20$ to 22\%, close to cross-encoder reranking of the same candidates (26\%) but without its inference cost. For topic modelling, we extend FASTopic with a seed-guided loss function and editorial abstract embeddings. Measurable coherence gains over the baseline remain minimal, but qualitative and temporal analysis shows that some topics track documented historical events. However, dominant topic weights barely exceed a uniform distribution, making topics unusable to inform retrieval. A pilot application of LLM-based concept induction (LLooM) further points towards a more human-interpretable alternative. Together, these results present a first approach to retrieving patristic references and modelling fine-grained topics in Bullinger's correspondence.
\end{abstract}




\section{Introduction}\label{sec:introduction}
For centuries, letters were the threads that wove together the intellectual world, enabling scholars to debate, collaborate, and exchange ideas across cities and countries, leaving behind a network of connections that shaped the history of thought \cite[p. 410]{edelstein2017historical}. One such record is the correspondence of Heinrich Bullinger (1504--1575), Swiss Reformer and Antistes of Zurich. The \textit{Heinrich Bullinger Briefwechsel} (HBBW), containing more than 13,000 surviving letters written predominantly in Latin and Early New High German (ENHG), documents religious, political, and intellectual exchanges in 16th-century Europe. Central to these exchanges are the negotiation and transmission of theological ideas in which Reformers frequently anchored their arguments in the authority of Scripture or the Church Fathers \parencite{bergjan_bullinger_2004, schindler2004bullinger}. One example is a letter written on November 1, 1534, from Martin Bucer (1491--1551) to Bullinger. In it, Bucer paraphrases Martin Luther's position that Christ's body is given bodily with, in, and under the sacramental symbols:

\begin{quote}
[...] \textit{cum se dominus symbolis offert, competere hic eas locutiones, quas scriptura et vulgatus usus in traditionibus vel commemoratione exhibitionum \textbf{rerum invisibilium} per \textbf{visibilia} \textbf{signa} usurpat, nempe ut tribuamus corpori domini, quae symbolis propria sunt.}\footnote{HBBW 4, Nr.~470, p.~391.\nocite{hbbw4}}

When the Lord offers himself through the symbols, those modes of expression are fitting here which Scripture and common usage employ in the handing-down or commemoration of exhibitions of invisible realities through visible signs; namely, that we attribute to the body of the Lord what properly belongs to the symbols.
\end{quote}

He uses close verbatim phrasing (see bolded words) from Augustine's \textit{De Catechizandis Rudibus} in which the sacrament's visible signs are described as symbols of the invisible reality they represent:

\begin{quote}

\textit{De sacramento [...] \textbf{signacula} quidem rerum divinarum esse \textbf{visibilia}, sed \textbf{res ipsas invisibiles} in eis honorari; }[...]

Concerning the sacrament [...] [he should be told] that visible things are indeed signs of divine realities, but that the invisible realities themselves are honoured in them.\footnote{Aug. \textit{catech. rud.} 26,50.}

\end{quote}

Bucer makes use of Augustinian claims to reframe Luther's real-presence doctrine in terms the Zurich Reformers, who held an Augustinian-inflected symbolic view, might accept. Recovering references like this requires recognising cases of intertextuality. Traditionally, detecting such cases relies on manual close reading, as has been done in the HBBW editions. Computational methods offer a complementary approach for the surfacing of potential intertextual relationships at scale. 

Two research gaps frame this paper. First, existing computational intertextuality detection in Latin has focused predominantly on biblical \cite{moritz_non-literal_2016, umphrey_investigating_2024, periti_trotr_2024, caffagni2025benchmarking} and classical \cite{Romanello_Citation_2016, berti2024digital, burns_profiling_2021, schelb_loci_2026} sources. The transfer of these methods to Reformation-era correspondence presents a distinct challenge as the Church Fathers' texts (circa 100--750 AD) and their 16th-century reuse are separated by roughly a millennium. Second, computational work on letter collections has emphasised the reconstruction of epistolary and intellectual networks as systems of actors and material exchange \cite{edelstein2017historical, mcshane2018visualising, poustie_letters_2014, ahnert_protestant_2015} with less attention to their discourse \cite{rossini_networked_2022, roller_tracing_2023}.

Addressing both gaps, this paper treats patristic references as a traceable signal of the theological arguments of the Reformers and topic structure as a proxy for the ideas they drew on or the context they were embedded in. It asks, \textbf{how can computational methods be used to detect patristic references and model theologically specific topics in Heinrich Bullinger's correspondence}? To answer this question, we investigate (RQ~1) the extent to which domain-adapted transformer embeddings can detect explicit and implicit patristic references in Bullinger's correspondence in comparison to a lexical baseline; and (RQ~2) the extent to which existing domain resources, i.e., thematic categories and letter abstracts from \textit{Bullinger Digital},\footnote{See \url{https://www.bullinger-digital.ch}, last accessed 14.08.2026} can guide topic models toward more coherent and interpretable topic structures. Together, these two objectives address complementary aspects of the same problem: the retrieval pipeline identifies \textit{which} patristic sources are referenced in a letter, while the topic model characterises \textit{what} is being discussed when they appear, thereby laying methodological groundwork for tracing how theological ideas propagated through the correspondence network.

\section{Related Work}\label{sec:related-work}
\subsection{Computational Detection of Text Reuse in Historical Sources}
Text reuse detection in historical documents is commonly formulated as an information retrieval (IR) task, in which candidate source passages are retrieved for a given text passage \parencite{moritz_non-literal_2016, manjavacas_feasibility_2019, caffagni2025benchmarking, schelb_loci_2026}.\footnote{One exception to this perspective is \textcite{burns_profiling_2021}, who conceptualise text reuse detection as a form of anomaly detection and claim that citations create stylistic breaks, vocabulary shifts, or a different syntax in texts.} Lexical retrieval methods retrieve passages based on shared terms, typically via inverted indices and ranking functions such as best-matching 25 (BM25) \parencite{hambarde2023information}. Such methods perform well for verbatim or lightly modified reuse but degrade in the presence of paraphrase and allusion \parencite{manjavacas_feasibility_2019}. Working with Jerome's letters, \textcite{schropp_towards_2024} show that even two-word overlaps can constitute intentional citation when validated through syntactic and contextual filtering. 

In dense retrieval, text is represented as continuous embeddings, enabling semantic matching with sentence encoders optimised for sentence-level similarity \parencite{hambarde2023information}. For Latin and Ancient Greek, dedicated token and sentence encoders were pretrained \parencite{riemenschneiderfrank-laberta, bamman_latin_2020, stroebel-roberta-base-latin-cased3, riemenschneider2023graecia} and fine-tuned \parencite{caffagni2025benchmarking, riemenschneider2023graecia, schelb_loci_2026, moritz_non-literal_2016}. \textcite{riemenschneider2023graecia} find token embeddings suboptimal for sentence similarity and therefore introduce SPhilBerta for cross-lingual retrieval. Similarly, \textcite{galatolo_dense_2023} show that a multilingual sentence encoder outperforms a Latin-specialised token model on semantic search.

First-stage retrieval, sparse or dense, rarely suffices alone and is typically followed by reranking or fusion of results \parencite{nogueira_multi-stage_2019, hambarde2023information, Bruch_2023}. Cross-encoder reranking, though computationally costly, generally improves ranking accuracy and generalisation over bi-encoders alone, including for Latin intertextuality \parencite{schelb_loci_2026}. 

Where prior work in Latin intertextuality has fine-tuned dense retrievers for the target domain \parencite{caffagni2025benchmarking, riemenschneider2023graecia, schelb_loci_2026}, we test how well models pretrained or fine-tuned for related intertextuality tasks transfer to Bullinger's correspondence. 
\subsection{Topic Modelling in Historical Sources}
\label{related-topicm}
Topic modelling aims to identify latent thematic structure in document collections, surfacing topics as ranked word sets or document clusters. A central challenge is interpretability, since without ground-truth labels topics are only useful insofar as they form human-coherent groupings \parencite{churchill_evolution_2022, chang2009reading}. 

Statistical approaches such as Latent Dirichlet Allocation (LDA) \parencite{blei2003latent} remain the field's baseline but struggle with multilingual, non-standardised historical sources \parencite{buning_topic_2019}.\footnote{Moreover, they lack a temporal dimension, which is addressed by subsequent variants \parencite{blei_dynamic_2006, wang2006topics}.} Bag-of-Words (BoW) models yield usable topic representations for historical (Latin) text but lack coherence \parencite{martinelli_exploring_2024, ginn_historia_2024}. Neural topic models instead leverage pretrained embeddings. BERTopic \parencite{grootendorst_bertopic_2022} provides coherence and yields usable topics on classical Latin \parencite{martinelli_exploring_2024} and Bullinger-era correspondence \parencite{strobel_decoding_2024}, but approximates document-topic assignments via cluster membership \parencite{ginn_historia_2024} and suffers from hyperparameter sensitivity and topic collapse \parencite{wu_survey_2024}. FASTopic \cite{wu2024fastopic} addresses this by embedding documents, topics, and vocabulary words in a shared semantic space, reducing the risk of topic collapse while retaining interpretable topic-word distributions.

To recover fine-grained domain-relevant themes that standard models dilute into dominant, coarse patterns \parencite{jeong_automatic_2025}, seed-guided approaches inject small sets of anchor keywords to bias topic discovery \parencite{jagarlamudi_incorporating_2012, zhang_effective_2023}, an approach shown to bridge computational output and domain-specific research questions in historical and social-science corpora \parencite{schnober_combining_2015, hurtado_bodell_seeded_2026}. 

Evaluation in the absence of ground truth relies primarily on coherence metrics \parencite{churchill_evolution_2022}, though these remain difficult to compare across methods \parencite{wu_survey_2024} and must ultimately be validated through expert qualitative review \parencite{martinelli_exploring_2024}.

More recently, approaches with large language models (LLMs) generate human-readable topic labels and keywords either directly or from statistically or neurally derived word distributions, such as TopicGPT \parencite{hankar_comprehensive_2025}. LLooM~\parencite{lam2024conceptinduction} is a method that synthesises sampled text passages, generates concepts based on pipeline generated criteria, applies those concepts to the data, and iteratively develops broader, higher-level abstractions with LLMs. 

\section{Data}\label{sec:data}
\subsection{Bullinger Digital Corpus (BDC)}
Over 13,000 letters related to Bullinger survive, critically edited since 1973 as the \textit{Heinrich Bullinger Briefwechsel}. The \textit{Bullinger Digital} project \cite{Scheurer_et_al_2025} digitised this correspondence and published it in TEI-XML with structured metadata (i.e., sender, recipient, date, place, language proportions, named-entity tags, thematic categories) on GitHub\footnote{\url{https://github.com/stazh/bullinger-korpus-tei}, last accessed 14.08.2026} and via an online database.\footnote{\url{https://www.bullinger-digital.ch/index.html}, last accessed 14.08.2026} Of the 13,114 catalogued letters, 1,256 lack a transcription, leaving a working corpus of 11,858 letters (hereafter \enquote{BDC}).

\subsection{Patristic Source Corpus (PSC)}
We collected works of 52 Church Fathers (700 files) from three digital repositories: the \textit{Corpus Corporum},\footnote{\url{https://mlat.uzh.ch} (CC), last accessed 14.08.2026} the \textit{Open Greek \& Latin Project},\footnote{\url{https://github.com/OpenGreekAndLatin} (OGL), last accessed 14.08.2026} and the \textit{Patristic Text Archive}.\footnote{\url{https://pta.bbaw.de/en/} (PTA), last accessed 14.08.2026} For a complete author and work list and the dataset itself, refer to the GitHub repository.\footnote{\url{https://github.com/len-rtz/patristic-sources}, last accessed 14.08.2026} Where available, we pooled Latin translations of Greek works alongside originals, as Bullinger's circle likely engaged with patristic texts in both languages or with Latin translations of Greek authors \cite[p. 133]{bergjan_bullinger_2004}. We deduplicated variant editions manually, preferring scholarly editions (e.g., \textit{Corpus Scriptorum Ecclesiasticorum Latinorum}) over the \textit{Patrologia Latina} (Migne\footnote{Jacques-Paul Migne, a priest, who published the works of the Church Fathers extensively during the 19th-century}). The pooled dataset may still be incomplete.\footnote{It reflects repository states at the time of data collection on 9th March 2026.} Due to inconsistent structural markup across repositories, we treated all sources uniformly by extracting all text nodes from \texttt{<body>} elements, which introduces editorial noise into some of the embeddings. Table \ref{tab:corpus-stats} reports corpus statistics for both PSC and BDC.

\begin{table}[h]
  \small
  \centering
  \begin{tabular}{@{}ll rrr@{}}
    \toprule
    & & \textbf{Files} & \textbf{Segments (1s)} & \textbf{Segments (3s)} \\
    \midrule
    \multirow{3}{*}{\textbf{BDC}}
      & Count        &     11,858 &     206,702 &     189,446 \\
      & Total tokens &  3,773,998 &   3,698,299 &  10,315,605 \\
      & Mean tokens  &        318 &          18 &          54 \\
    \addlinespace
    \multirow{3}{*}{\textbf{PSC}}
      & Count        &        700 &     832,556 &     831,172 \\
      & Total tokens & 20,949,791 &  16,321,764 &  48,867,504 \\
      & Mean tokens  &     29,928 &          20 &          59 \\
    \bottomrule
  \end{tabular}
  \caption{Corpus statistics for BDC and PSC: full source files, single-sentence segments (1s), and 3-sentence overlapping window segments (3s). Segments shorter than 4 tokens were dropped during chunking, which is reflected in total tokens of segments.}
  \label{tab:corpus-stats}
\end{table}

\subsection{Retrieval Evaluation Dataset}
\label{sec:evaluationdata}
We created a dataset of annotated patristic references to evaluate whether dense retrieval finds explicit (cf.~Figure \ref{fig:verbatim}) and implicit patristic references (cf.~Figures \ref{fig:theo-allusion} and \ref{fig:trans-allusion}). We define explicit references as verbatim quotations, and implicit references as all other forms. The dataset comprises 100 annotations, 50 explicit and 50 implicit, across 67 letters spanning 19 Church Fathers dominated by Augustine (54\%), Jerome (12\%), and Tertullian (9\%). We identified these ground truth references by extracting scholarly edited footnotes from the HBBW in proximity to XML-TEI person-tags of Church Fathers, through string alignment detection using Passim~\parencite{smith_passim} against a scraped CC dataset for explicit citations, and close reading of footnotes in digitally published HBBW letter editions.\footnote{\url{https://www.tvz-verlag.ch/reihe/heinrich-bullinger-werke-8/?page_id=1}, last accessed 14.08.2026} Because annotation relied on established scholarly identifications for implicit references and direct textual correspondence for explicit ones, a single annotator annotated the letters using INCEpTION \parencite{klie-etal-2018-inception}. The full methodology for constructing this dataset and the dataset itself can be found on GitHub.\footnote{\url{https://github.com/len-rtz/bullinger-patristic-annotations}, last accessed 14.08.2026} The 67 letters were not exhaustively annotated, meaning not every known patristic reference in a letter is marked. This does not directly affect our retrieval evaluation, which assesses whether the system retrieves the source text associated with each annotated reference, but limits the dataset’s reuse for other purposes, e.g. as training data. The dataset is predominantly Latin-to-Latin, with five Greek-to-Latin and four Latin-to-ENHG cross-lingual annotations, which limits multilingual generalisability while mirroring the BDC's language distribution of roughly 80\% Latin. A Jaccard analysis of lemmatised Latin annotations shows explicit references average 70\% lexical overlap with source texts, versus 12\% (max 31\%) for implicit ones.

\begin{figure}[h!]
\centering
{\scriptsize
\MyBox{myred}{Explicit Verbatim}
{Nolo nobis credatur, scripturę recitentur. \\ \newline \textit{Bullinger Digital, HBBW 1, Nr.~16, p. ~100\nocite{hbbw1}}}
{Nolo nobis credatur, Scriptura recitetur. \\ \newline Ambr.~ \textit{incarn.}~3,14}
{Do not let it be believed on our authority; let the Scriptures be recited.}
{}}
\caption[{Example of an Explicit Reference.}]{Example of an explicit reference and its translation.}
\label{fig:verbatim}
\end{figure}

\begin{figure}[h!]
\centering
{\scriptsize
\MyBox{myred}{Implicit Theological Allusion}
{Rursus, cum sacramenta sint verba visibilia, quae docent sensibiliter, nomenque ipsum, quorum symbola sunt, sortiantur, nihil offendit nos, cum audimus panem dici corpus Christi. \\ \newline \textit{Bullinger Digital, HBBW 4, Nr.~482, p.~421}}
{Accedit verbum ad elementum, et fit Sacramentum, etiam ipsum tanquam visibile verbum. \\ \newline Aug.~\textit{in Joh.~ tract.}~80,3}
{Again, since \textbf{sacraments are visible words}, which teach through the senses, and since they take their name from the things of which they are symbols, nothing should offend us when we hear bread being called the body of Christ.}
{The \textbf{word is added to the element, and it becomes a sacrament}, itself also being, as it were, \textbf{a visible word}.}}
\caption[Example of an Implicit Theological Reference.]{Example of an implicit theological reference and its translation.}
\label{fig:theo-allusion}
\end{figure}

\begin{figure}[h!]
\centering
{\scriptsize
\MyBox{myred}{Implicit Metaphorical Allusion}
{Scribere tamen distuli, ne ferventi adhuc animo et inopinata provocatione quorundam commoto nimium crederem; \textbf{iacula etenim non provisa altius feriunt.} \\ \newline \textit{Bullinger Digital, HBBW 10A, Nr.~372a, p.~46\nocite{hbbw10a}}}
{\textbf{Minus enim jacula feriunt quae praevidentur;} et nos tolerabilius mundi mala suscipimus, si contra haec per praescientiae clypeum munimur. \\ \newline Greg.~M. \textit{hom.~in ev.}~35,1}
{Nevertheless, I delayed writing, with my mind still heated and stirred by the unexpected provocation of certain people, I might judge too harshly; \textbf{for missiles that are not foreseen strike more deeply.}}
{\textbf{For missiles that are foreseen wound less severely;} and we endure the evils of the world more tolerably if we are armed against them with the shield of foreknowledge.}}
\caption[Example of an Implicit Metaphorical Reference.]{Example of an implicit metaphorical reference.}
\label{fig:trans-allusion}
\end{figure}

\section{Methodology}
\subsection{Dense Retrieval of Patristic References}
Following \textcite{allenbach1975bibliapatristica}, we can define patristic text reuse as a textual unit within a letter in which the author \enquote{refers, explicitly or implicitly, through quotation, mention or reminiscence, to one or more parts} (p.~20) of patristic writings. Within such a unit, the writer’s argument and the patristic material can be interwoven, with verbal overlap potentially being minimal or entirely absent. At the same time, even a single term may constitute a patristic reference and therefore form a reuse unit~\cite{hue-gay_tei-encoding_2017}. 

When patristic text reuse detection is framed as an IR task~\cite{moritz_non-literal_2016, manjavacas_feasibility_2019, caffagni2025benchmarking, schelb_loci_2026}, we utilise text segments from the BDC as query documents \textit{q} and text segments from sources in the PSC as candidate documents \textit{d}. The IR task objective is retrieving the most semantically similar documents for each query, ranked by a relevance score \textit{s(q,d)}.

For both corpora, we combine sentence-level segments \cite{caffagni2025benchmarking, schelb_loci_2026} with 3-sentence overlapping windows to capture references spanning sentence boundaries (cf.~Table \ref{tab:corpus-stats}). Fixed windows were rejected as they do not align with semantic units \cite{manjavacas_feasibility_2019, schmidt_sui_2024}. 

We evaluate three bi-encoders, one monolingual and two multilingual, each available as a base model and as a variant fine-tuned for (Latin) intertextuality. This lets us isolate how far task-specific fine-tuning transfers to a new domain. The base models are \nolinkurl{LaBERTa}~\cite{riemenschneiderfrank-laberta}; \nolinkurl{SPhilBerta}~\cite{riemenschneider2023graecia}, a cross-lingual sentence encoder for Latin, Ancient Greek, and English; and modern multilingual text encoder \nolinkurl{multilingual-e5-large}~\cite{wang2024multilingual}, previously applied to the Bullinger corpus~\cite{strobel_decoding_2024}. Their fine-tuned variants are \nolinkurl{LaBERTa-W\_VULG-S\_VL}~\cite{caffagni2025benchmarking} (hereafter \nolinkurl{LaBERTa-VL}), chosen despite its monolingual scope because it was trained on the related task of biblical reuse in patristic texts, and the LociSimiles models~\cite{schelb_loci_2026}.

Given the query set $Q$ and candidate database $D$, the three models produce embeddings $\bar{q}= f_{\theta}(q) \in \mathbb{R}^{768/1024} \quad \text{for } q \in Q$ and $\bar{d}_i= f_{\theta}(d_i) \in \mathbb{R}^{768} \quad \text{for } d_i \in D$ where $f_{\theta}$ represents the respective encoder. Cosine similarity quantifies query--candidate relevance.

Bi-encoder similarity search uses FAISS;\footnote{\url{https://faiss.ai/index.html}, last accessed 14.08.2026} the LociSimiles pipeline handles candidate generation internally via ChromaDB.
For cross-encoder reranking, we use the LociSimiles classifier \nolinkurl{PhilBerta-class-latin-intertext-v2}~\parencite{schelb_loci_2026}, which jointly encodes query-candidate pairs and classifies each as intertextually related or not, requiring one forward pass per candidate. In two-stage pipelines, the bi-encoder retrieves the top-100 candidates per query, which the cross-encoder then reranks.

\begin{figure}
    \centering
    \includegraphics[width=\textwidth]{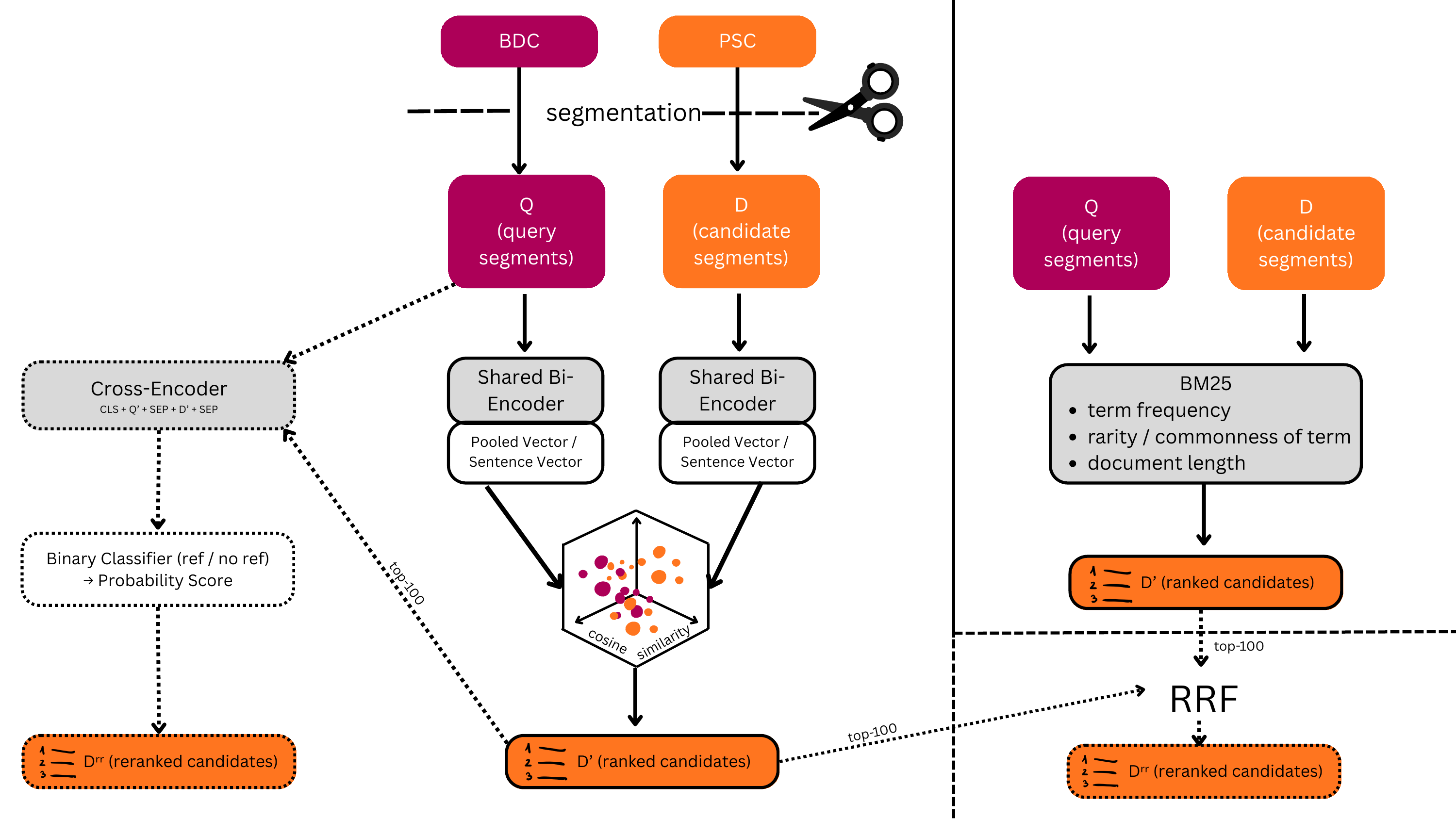}
    \caption{Patristic intertextuality retrieval pipeline. Left: For bi-encoder retrieval, BDC query segments $Q$ are encoded alongside PSC candidate segments $D$. Cosine similarity retrieval produces a ranked list $D'$; in a variation the top-100 candidate documents are reranked by a cross-encoder binary classifier. Right: BM25 computes relevance scores between $Q$ and $D$ for ranking. In a variation, the top-100 candidates from both dense and sparse retrieval are fused using RRF. Dashed borders indicate variational components of the pipeline.}
    \label{fig:metho-1}
\end{figure}

We evaluate on the evaluation dataset using $Recall@k$ and mean reciprocal rank ($MRR$), stratified by reference type (explicit / implicit), and compare against BM25 as a lexical baseline and a hybrid setup using RRF (cf.~Figure \ref{fig:metho-1}).

\subsection{Topic Modelling}
For topic modelling, we use FASTopic \cite{wu2024fastopic}, which derives topic-word and document-topic distributions from a shared embedding space while retaining an interpretable BoW representation.\footnote{Its advantages over LDA (coherence) and BERTopic (stable document-topic assignments) were discussed in Section \ref{related-topicm}.}
We restrict BoW input to POS-tagged nouns and proper nouns \parencite{wu2024fastopic, strobel_decoding_2024}, lemmatised and lowercased with LatinCy (\nolinkurl{la_core_web_lg}) \parencite{burns2023latincy} for Latin, spaCy (\nolinkurl{de_core_news_lg}) for German (regests) and ENHGTagger\footnote{\url{https://www.cis.uni-muenchen.de/~schmid/tools/HistoricalGermanTools/}, last accessed 14.08.2026} for ENHG. We exclude tokens tagged \nolinkurl{<persName>} or \nolinkurl{<placeName>} during TEI-XML extraction and save them to a stopword list, extended by manually curated high-frequency terms identified through iterative inspection of topic outputs.\footnote{The stopword list is available in the topic modelling repository \url{https://github.com/len-rtz/bullinger-topic-modelling/blob/main/data/stopwords_full.json}, last accessed 14.08.2026} We drop terms appearing in fewer than ten documents, yielding an effective vocabulary of 8,316 unique nouns. Document representations use the multilingual bi-encoder \nolinkurl{multilingual-e5-large-emb-lat-intertext-v1}.

\subsubsection{Category-Guided Topic Initialisation (Seeded-FASTopic)}
The existing eleven \textit{Bullinger Digital} categories\footnote{The categories result from an annotation combining expert-defined category sets, manual annotation of a sample, and LLM prompting, \url{https://github.com/bullinger-digital/bullinger-auxiliary-scripts/tree/main}, last accessed 14.08.2026} flatten internal distinctions. For example, \enquote{Faith and Doctrine in Theory} conflates biblical interpretation, church discipline, and the Eucharist controversy. FASTopic normally initialises $k$ topic embeddings randomly. We inject the corpus's known thematic structure by expanding the categories into 16 seed topics, splitting heterogeneous categories into finer-grained distinctions and directly setting the minimum $k$ (cf. Appendix \ref{append:seedtopics}). Following \textcite{zhang_effective_2023}, we extract meaning-bearing terms from each category's English description (e.g., \enquote{martyrdom}, \enquote{Eucharist controversy}, \enquote{military action}), embed them with the multilingual encoder, and match each to its top-10 nearest neighbours in the vocabulary by cosine similarity. Each seed-assigned topic $k$'s initial topic embedding is the centroid $\bar{e}_{S_k}$ of its matched vocabulary embeddings $e_w \in \mathbb{R}^H$, with $S_k$ being the induced seed vocabulary. We add a regularisation term to the objective to keep these embeddings influential during training:
\begin{equation}
    \mathcal{L}_\text{seed} = \frac{1}{|\mathcal{K}_s|} \sum_{k \in \mathcal{K}_s} 
    \left\| T_k - \bar{e}_{S_k} \right\|^2
\end{equation}
where $\mathcal{K}_s \subseteq \{1, \ldots, K\}$ denotes the set of seed-assigned topics. Since the centroids $\bar{e}_{S_k}$ and topics $T_k$ are both outputs of the same encoder, they are directly comparable. The loss function penalises training by the average squared Euclidean distance between the seeded topic embedding at a given training step and its centroid.

The full training objective of the altered FASTopic algorithm becomes $
    \mathcal{L} = \mathcal{L}_\text{DSR} + \mathcal{L}_\text{ETP} 
    + \mathcal{L}_\text{seed}$
We refer to this approach as Seeded-FASTopic. Setting $\mathcal{L}_\text{seed} = 0$ recovers standard FASTopic. The altered algorithm can be found in the GitHub repository\footnote{\url{https://github.com/len-rtz/bullinger-topic-modelling}, last accessed 14.08.2026}.

\subsubsection{Regest Embeddings}
Regests (editorial abstracts from the HBBW critical editions) are incorporated into the embedding process where available, following \textcite{strobel_decoding_2024}'s proposal that this could provide a stronger thematic signal than the letter text alone.

The letter-regest embedding $d_i$ is constructed as a weighted interpolation: $ d_i = \alpha \cdot f_\text{doc}(x_i^\text{letter}) + (1 - \alpha) \cdot f_\text{doc}(x_i^\text{regest})$ where $\alpha \in [0, 1]$ controls the relative weight of letter versus editorial content. For letters without abstracts, $\alpha = 1$. The German abstracts from the HBBW critical editions are currently available for 3,233 letters (27.3\%), covering correspondence up to the 1540s.

\begin{figure}[h!]
    \includegraphics[width=\textwidth]{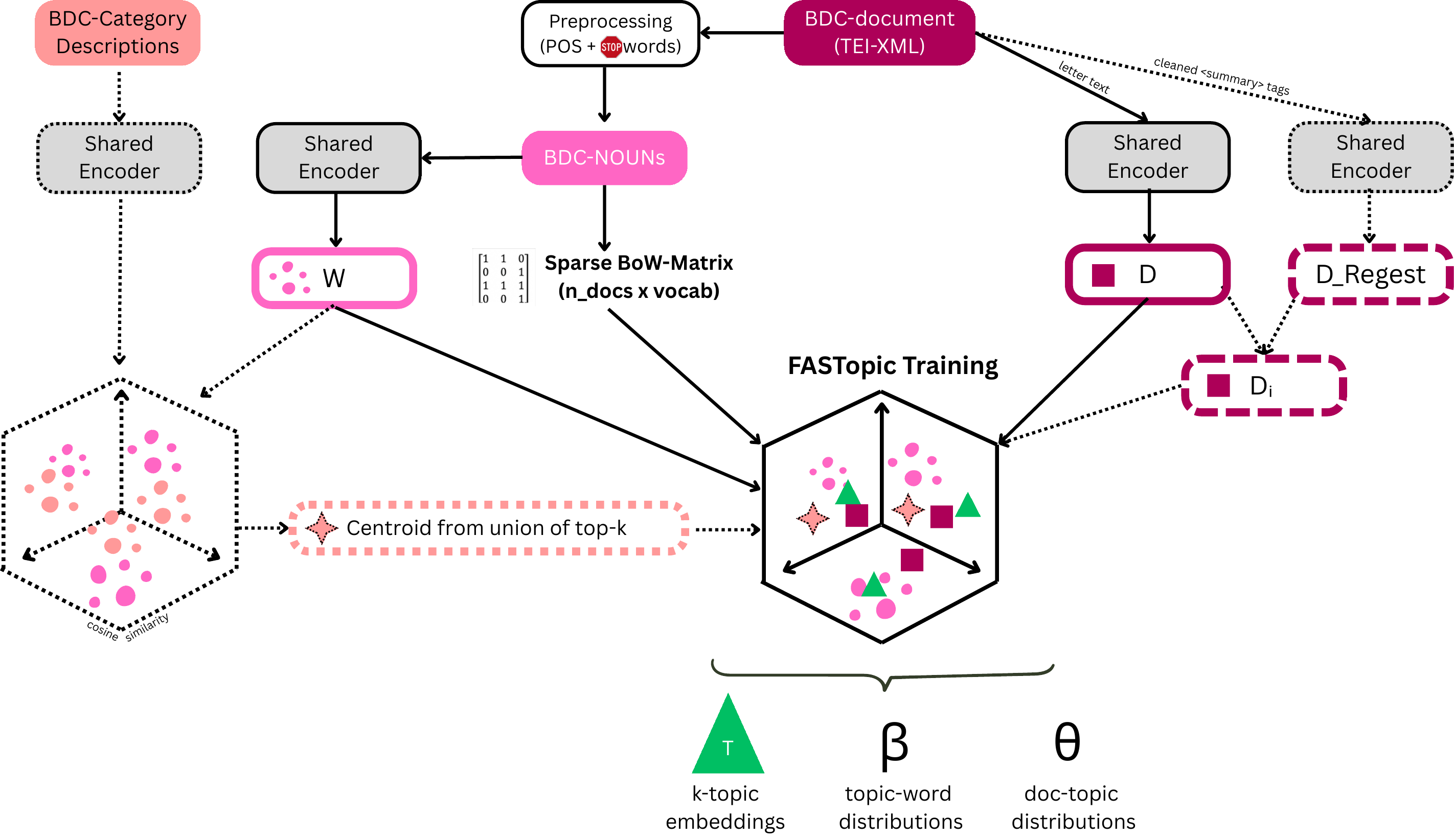}
    \caption[Topic Modelling Pipeline]{(Seeded-)FASTopic training pipeline (with regest augmentation). BDC documents are POS-tagged and reduced to noun-only (BDC-NOUNs), from which a BoW matrix and vocabulary embeddings $W$ are derived. In the seeded setup, category description nouns are embedded and matched to $W$ via cosine similarity, with top-10 neighbours averaged into seed centroids initialising 16 topic embeddings. Document embeddings $D$ are optionally interpolated with regest embeddings. FASTopic produces topic embeddings $T$, topic-word distributions $\beta$, and document-topic distributions $\theta$. Dashed borders indicate variational components.}
    \label{fig:metho-2}
\end{figure}

FASTopic is applied to the full BDC corpus across the configurations \textit{baseline, Seeded, Regest}, and \textit{Seeded+Regest}. All configurations use \texttt{DT\_alpha} = 15 and \texttt{min\_df} = 10\footnote{Recommended by \textcite{wu2024fastopic}; see also \url{https://pypi.org/project/fastopic/} (last accessed 14.08.2026).}. $\alpha$, the interpolation weight is set to 0.5, assigning equal weight to the original and regest representations. Since regests are available for only 27.3\% of letters, this introduces a structural asymmetry. Letters with regests receive a blended representation while the remainder rely on letter text alone, a potential confound that future work should address through ablation over $\alpha$ values. Figure \ref{fig:metho-2} presents the method.

Topic quality is assessed with topic diversity (proportion of unique words across the top-$n$ words per topic) and topic coherence (measuring co-occurrence of top words against an external reference corpus\footnote{The \textit{Theologenbriefwechsel im Südwesten des Reichs in der Frühen Neuzeit} (ThBW) \parencite{thbw2020}.}), following \textcite{wu2024fastopic}. All configurations are evaluated across five random initialisations; qualitative analysis is based on a single run per configuration. BERTopic with the same encoder serves as an additional comparison.

\section{Results and Evaluation}
\subsection{Dense Retrieval of Patristic References}
Table \ref{tab:model-comparison} presents the performance of the selected bi-encoders, both standalone and with cross-encoder reranking, on the evaluation dataset. Among the standalone bi-encoder retrieval configurations, \nolinkurl{LaBERTa-VL}, with mean-pooling token embeddings, achieves the highest overall $Recall@k$ ($R@k$) across all cutoffs (42--49\%), outperforming its base model (35--40\%), demonstrating that \textcite{caffagni2025benchmarking}'s biblical fine-tuning transfers moderately to patristic references in Reformation correspondence. Table \ref{tab:sim-score-laberta} shows a selection of query-candidate pairs and their cosine scoring. Notably, it also achieves the strongest implicit recall among all standalone bi-encoders (6--16\% $R@k$), while every other bi-encoder struggles to exceed 2--6\% implicit recall. 
The fine-tuned models by \textcite{schelb_loci_2026} perform better than their base models almost at every cutoff, though gains concentrate in explicit recall rather than implicit detection. Especially \nolinkurl{SPhilBerta} in its original form performs poorly as a standalone retriever (overall 12\% $R@k$ across all cutoffs); its fine-tuned version \parencite{schelb_loci_2026} improves on overall recall (25--34\% $R@k$). As the second section of Table \ref{tab:model-comparison} shows, embedded in the two-stage retrieval pipeline, its performance improves ranking from 34\% to 41\% $R@k20$, suggesting that the LociSimiles \nolinkurl{PhilBerta-class-latin-intertext-v2} classifier adds real signal on top of a weaker base retriever. Combining \nolinkurl{LaBERTa-VL} with the classifier also improves implicit metrics at higher rank cutoffs. The clearest gains are in implicit reference detection, as $R@5$ more than doubles from 6\% to 14\%. 

The results should be by noting, that the evaluation dataset includes nine cross-lingual references which the Latin encoders are not designed for. Only one was retrieved by \nolinkurl{mE5-LS}. The reported scores therefore primarily reflect Latin-to-Latin retrieval performance.

We situate BM25 as a lexical baseline against the neural retrieval approaches. As can be seen in Table \ref{tab:model-comparison}, it outperforms all bi-encoder configurations on explicit references (88--94\% $R@k$). This reflects the structural advantage of verbatim alignment for explicit quotation. The competitive numbers for implicit recall (12--16\%, $R@k$) of the lexical approach warrant closer inspection. The implicit references detected by BM25 are cases with short shared-vocabulary paraphrasing. This is where the bi-encoder falls short as its embeddings dilute short paraphrase among the surrounding words. 
For example, BM25 correctly matches a query from letter HBBW 7, Nr.~980, (p.~121)\nocite{hbbw7}: \enquote{\textit{\textbf{Det illis Deus} mentem saniorem et vobis in bono \textbf{perseverantiam}}}\footnote{\textit{May God give them a sounder mind, and grant you perseverance in what is good.}} to the passage from Augustine's \textit{De correptione et gratia} with lexical overlap, namely \enquote{\textit{...ejus qui non \textbf{perseveraverit}...}} and \enquote{\textit{...ne forte \textbf{det illi Deus} poenitentiam...}}\footnote{Full quote: \enquote{Sed non ideo est ejus qui non perseveraverit, negligenda correptio, ne forte det illi Deus poenitentiam, et resipiscat de diaboli laqueis.} Aug.~\textit{corrept.}~6,10. \textit{But for this reason the rebuke of one who has not persevered is not to be neglected — lest perhaps God give him repentance, and he come to his senses again, free from the snares of the devil.}}. One example for a match that BM25 misses but bi-encoder retrieval finds, is a segment from letter HBBW 4, Nr.~466 (p. 377) \nocite{hbbw4}: 

\begin{quote} {\textit{Hinc vulgo illud pertritumc iactari: «Sacramentum est invisibilis gratiae visibile signum.» Visibilia ergo signa in coena esse panem et vinum, at invisibilia redemptionem per Christum partam, corpus et sanguinem domini, quae fidelibus a solo Christo divina offerantur sive exhibeantur virtute, a ministro autem ministerialiter et a sanctis mente, percipiantur spiritualiter.}}\footnote{\textit{Hence that well-known and oft-repeated saying is commonly quoted: «A sacrament is a visible sign of invisible grace.» Accordingly, the visible signs in the Lord’s Supper are bread and wine, whereas the invisible realities are the redemption obtained through Christ, and the body and blood of the Lord. These are offered or presented to the faithful by Christ alone through divine power, by the minister in a ministerial capacity, and are received spiritually by the saints through the mind.}} \end{quote} 

which \nolinkurl{LaBERTa-VL} correctly links to a passage from Augustine's \textit{De catechizandis rudibus}: \begin{quote}
{\textit{De sacramento sane quod accipit, cum ei bene commendatum fuerit, signacula quidem rerum divinarum esse visibilia, sed res ipsas invisibiles in eis honorari; nec sic habendam esse illam speciem benedictione sanctificatam, quemadmodum habetur in usu quolibet: dicendum etiam quid significet et sermo ille quem audivit, quid in illo condiat, cujus illa res similitudinem gerit.}}\footnote{Aug.~\textit{catech.~rud.}~26,50. \textit{On the subject of the sacrament, indeed, which he receives, it is first to be well impressed upon his notice that the signs of divine things are, it is true, things visible, but that the invisible things themselves are also honoured in them, and that that species, which is then sanctified by the blessing, is therefore not to be regarded merely in the way in which it is regarded in any common use.}}    
\end{quote}

While the passages share some lexical elements relating to visibility and invisibility, they extend beyond vocabulary to a semantic parallel. Bullinger, author of the letter, engages with Augustine's formulation by applying the visible sign/invisible reality distinction specifically to the Eucharist. Comparing the retrieval results, the dense and sparse approaches recover disjoint implicit reference sets. Counting a query as retrieved if either method, e.g., \nolinkurl{BM25} and \nolinkurl{LaBERTa-VL}, finds it, could therefore achieve increased implicit recall, which motivates a dual retrieval strategy for implicit reference detection.

Using reciprocal rank fusion RRF \parencite{cormack2009reciprocal}, which merges retrieval results with different scoring scales by combining their rankings, we fuse retrieval rankings from the best-performing bi-encoder \nolinkurl{LaBERTa-VL} with the BM25 ranking. The last section in Table \ref{tab:model-comparison} reports its results. The fusion improves overall recall relative to either method alone. For comparison we pass the same hybrid candidate pool through the cross-encoder classifier. At $R@20$, the classifier pipeline reaches slightly higher implicit recall (26.0\% vs. 22.0\% for RRF-pipeline) but matches the RRF retrieval at smaller cutoffs. The RRF pipeline achieves higher explicit recall at every cutoff (90.0--94.0\% vs. 62.0--84.0\%) suggesting that the classifier favours apparent semantic similarity over lexical overlap. Given that the classifier adds computational costs for the hybrid pool at scale, RRF positions itself as a more favourable cost-performance option.

\begin{table}[]
    \centering
    \small
    \begin{tabular}{p{6cm}|p{6cm}|p{1cm}}
    \toprule
         \textbf{Query Text and ID} & \textbf{Candidate Text and ID} & \textbf{Score}\\
         \midrule
         Quia non potestis audire sermonem meum. \newline {\tiny 12080\_sent\_277} & Quia non potestis audire sermonem meum. \newline {\tiny 035\_Augustinus-Hipponensis\_In-Joannis-evangelium-tractatus-CXXIV\_sent\_8501 }& 1.00 \\
         \midrule
         128) apertissime hoc defendit et ita loquitur: Unus et unus deus, deus ingenitus, deus unigenitus. De patre et filio, non duo dii, inquit [...] \newline {\tiny 5473\_sent\_51\_53}& Ita in Deo patre et in Deo filio neque duos connumerabis deos, quia unum uterque sunt; neque singularem praedicabis, quia uterque non unus [...] \newline {\tiny 010\_Hilarius-Pictaviensis\_De-Trinitate\_sent\_2541\_2543}& 0.80 \\
         \midrule
         Ut enim patri simpliciter nomen dei tribui videmus in scripturis, ita filio nunquam dari nisi aperta mentione nominis eius, vel verbi [...] \newline {\tiny 5473\_sent\_26\_28 } & Proprium Dei filio est, ne praeter eum Deus sit; proprium Deo patri est, ne absque eo Deus quisquam sit: verbis Dei de Deo utere. \newline {\tiny 010\_Hilarius-Pictaviensis\_De-Trinitate\_sent\_1233} & 0.72 \\
         \midrule
         Quo magis miror, cum vestra lego, quid tam diserte affirmetis panem id, quod est, manere, cum id Lutherus nunquam negarit, imo primus affir [...] \newline {\tiny 10506\_sent\_11\_13} & Proinde quod non ita fit, quamvis sit panis et calix, alimentum est refectionis, non sacramentum religionis: nisi quod benedicimus, gratias [...]\newline {\tiny 042\_Augustinus-Hipponensis\_Contra-Faustum-Manichaeum\_sent\_2381\_2383} & 0.67 \\
         \midrule
         Quis tam scelestus, tam nepharius tamque impius? Quae lingua tam virulenta ac lȩtifera? Disperdat dominus universa labia dolosa! \newline {\tiny 10015\_sent\_6\_8} & Ream linguam non facit, nisi mens rea. Quis est autem qui non fallatur, etsi noluit fallere? Quis est homo cui non subrepat fallacia? {\tiny 038\_Augustinus-Hipponensis\_Sermones-de-Scripturis\_sent\_27997\_27999}& 0.56 \\
         \midrule
         Oportet enim nos implere omnem iustitiam. \newline {\tiny 12805\_sent\_144} & Dico enim vobis, quia nisi abundaverit justitia vestra plus quam Scribarum et Pharisaeorum, non intrabitis in regnum coelorum: id est, nisi [...] \newline {\tiny 034\_Augustinus-Hipponensis\_De-sermone-Domini-in-monte\_sent\_142} & 0.43 \\
         \bottomrule
    \end{tabular}
    \caption{Exemplary \nolinkurl{LaBERTa-VL} retrieval results across cosine similarity range, from an exact match (score = 1.00) to a low-scoring pair (score = 0.43).}
    \label{tab:sim-score-laberta}
\end{table}

\begin{table}[]
\small
\centering
\begin{tabular}{llcccc}
\toprule
\textbf{Bi-Encoder} & & \textbf{R@5} & \textbf{R@10} & \textbf{R@20} & \textbf{MRR@5} \\
\midrule
\multirow{3}{*}{LaBERTa}
  &   & 35.0 & 38.0 & 40.0 & 0.23 \\
  & Explicit & 68.0 & 72.0 & 76.0 & 0.46 \\
  & Implicit & 2.0 &  4.0 &  4.0 &  0.01 \\
\midrule
\multirow{3}{*}{\shortstack[l]{%
LaBERTa-W\_VULG-S\_VL\\
{\tiny(LaBERTA-VL) \parencite{caffagni2025benchmarking}}}}
  &   & \textbf{42.0} & \textbf{45.0} & \textbf{49.0} & \textbf{0.31} \\
  & Explicit & \textbf{78.0} & \textbf{80.0} & \textbf{82.0} & \textbf{0.58} \\
  & Implicit &  \textbf{6.0} &  \textbf{10.0} &  \textbf{16.0} &  \textbf{0.03} \\
\midrule
\multirow{3}{*}{multilingual-e5-large}
  &   & 27.0 & 30.0 & 32.0 & 0.20 \\
  & Explicit & 52.0 & 58.0 & 62.0 & 0.40 \\
  & Implicit &  2.0 &  2.0 &  2.0 &  0.01 \\
\midrule
\multirow{3}{*}{\shortstack[l]{%
multilingual-e5-large-emb-lat-intertext-v1 \\ {\tiny (m-e5-LS) \parencite{schelb_loci_2026}}}}
  &   & 35.0  & 36.0 & 36.0 & 0.25 \\
  & Explicit  & 70.0 & 70.0 & 70.0 & 0.50 \\
  & Implicit  & 0.0 & 2.0 & 2.0 & 0.00 \\
\midrule
\multirow{3}{*}{\shortstack[l]{%
SPhilBerta\\ 
{\tiny \parencite{riemenschneider2023graecia}}}}
  &   & 12.0 & 12.0 & 12.0 & 0.11 \\
  & Explicit & 22.0 & 22.0 & 22.0 & 0.20 \\
  & Implicit &  2.0 &  2.0 &  2.0 &  0.02 \\
\midrule
\multirow{3}{*}{\shortstack[l]{%
SPhilBerta-emb-lat-intertext-v1 \\ {\tiny (SPhilBerta-LS) \parencite{schelb_loci_2026}}}}
  &          & 25.0 & 30.0 & 34.0 & 0.20 \\
  & Explicit & 48.0 & 58.0 & 62.0 & 0.36 \\
  & Implicit &  2.0 &  2.0 &  6.0 &  0.02 \\
\midrule
\textbf{Retrieve-and-Rerank Pipeline} & & &  &  &  \\
\midrule
\multirow{3}{*}{\shortstack[l]{%
LaBERTa-W\_VULG-S\_VL\\
+ PhilBerta-class-latin-intertext-v2}}
  &   & 37.0 & \textbf{46.0} & \textbf{52.0} & 0.24 \\
  & Explicit & 60.0 & 76.0 & \textbf{82.0} & 0.40\\
  & Implicit &  \textbf{14.0} &  \textbf{16.0} &  \textbf{22.0} &  \textbf{0.08} \\
\midrule
\multirow{3}{*}{\shortstack[l]{%
LociSimiles e5\\
{\tiny \parencite{schelb_loci_2026}}}}
  &   & 35.0 & 41.0 & 43.0 & \textbf{0.29} \\
  & Explicit & 70.0 & \textbf{78.0} & 80.0 & \textbf{0.58} \\
  & Implicit &  0.0 &  4.0 &  6.0 &  \ 0.00 \\
\midrule
\multirow{3}{*}{\shortstack[l]{%
LociSimiles SPhilBerta\\
{\tiny \parencite{schelb_loci_2026}}}}
  &   & \textbf{38.0} & 40.0 & 41.0 & 0.27 \\
  & Explicit & 68.0 & 72.0 & 74.0 & 0.50 \\
  & Implicit & 8.0 & 8.0 &  8.0 &  0.04 \\
\midrule
\textbf{Lexical} & & &  & &  \\
\midrule
\multirow{3}{*}{BM25}
  &  & \textbf{50.0} & \textbf{53.0} & \textbf{55.0} & \textbf{0.39}\\
  & Explicit & \textbf{88.0} & \textbf{92.0} & \textbf{94.0} & \textbf{0.69} \\
  & Implicit &  12.0 &  14.0 &  16.0 &  \textbf{0.10} \\
\midrule
\textbf{Hybrid} & & &  & &  \\
\midrule
\multirow{3}{*}{\shortstack[l]{%
LaBERTa-W\_VULG-S\_VL\\
+ BM25\\
RRF-used}}
  &   & \textbf{52.0} & \textbf{56.0} & \textbf{58.0} & \textbf{0.37} \\
  & Explicit & \textbf{90.0} & \textbf{94.0} & \textbf{94.0} & \textbf{0.64}\\
  & Implicit &  \textbf{14.0} &  \textbf{18.0} & 22.0 &  \textbf{0.09} \\
\midrule
\multirow{3}{*}{\shortstack[l]{%
LaBERTa-W\_VULG-S\_VL\\
+ BM25 \\
+ PhilBerta-class-latin-intertext-v2}}
  &   & 38.0 & 47.0 & 55.0 & 0.25 \\
  & Explicit & 62.0 & 76.0 & 84.0 & 0.44\\
  & Implicit &  \textbf{14.0} &  \textbf{18.0} & \textbf{26.0} &  0.07 \\
\bottomrule
\end{tabular}
\caption[Performance of detection on evaluation dataset by reference type, bi-encoder]{Performance comparison on evaluation dataset (n=100) in percentage. $R@k$=Recall at rank $k$ and $MRR@k$=Mean Reciprocal Rank at rank $k$. Best values, per setup, are shown in bold.}
\label{tab:model-comparison}
\end{table}

\subsection{Topic Modelling}
As Table \ref{tab:stability} shows, mean coherence ($C_V$) peaks at $k = 16$ for the baseline and seeded approach with 0.420 across all FASTopic configurations. Topic diversity (TD) remains between 0.937 and 1.0 across all configurations and $k$, indicating distinct topic vocabularies. Both degrade as $k$ increases. An analysis of dominant topic weights across the letters (cf. Table~\ref{tab:topic-distribution}) shows that as $k$ increases, mean dominant weights decrease monotonically. The numbers reflect simply a redistribution of probability mass across a larger topic inventory. Neither seeding nor the inclusion of regests alters the overall diffuseness of the distributions. We initially hypothesised that topic distributions could enhance retrieval, assuming that shared references cluster within the same topics. The dominant topic weights argue against this: at $k = 20$, a uniform distribution would assign each topic 0.05, while the mean dominant weight reaches 0.062--0.066 (Table~\ref{tab:topic-distribution}). Since the strongest topic of a letter is thus barely distinguishable from the rest, we did not pursue topic-based filtering or reranking in this study.

As a further comparison, BERTopic with HDBSCAN runs yield the highest mean $C_V$ (0.475) but only a mean topic diversity of 0.623. It exhibits severe run-to-run instability, with data-driven discovered topic count varying from 3 to 22 across five UMAP seeds, suggesting that density-based clustering in this embedding space does not produce a reliable or reproducible topic structure. 

All FASTopic configurations recover semantically distinct clusters spanning topics such as theological discourse, military affairs, domestic matters, and print culture. Increasing the number $k$ comes at the cost of topic duplication and increased noise across all runs (cf.TD in Table \ref{tab:stability}). Beyond this, topic vocabulary varies considerably across runs (see Table \ref{tab:stability}). The same thematic space is partitioned differently depending on $k$ and run, with words shifting between topics and appearing or disappearing across configurations. The minimal differences in quantitative metrics across conditions at matched $k$ values motivate a qualitative inspection. For all outputs of the here discussed runs, refer to the repository.\footnote{\url{https://github.com/len-rtz/bullinger-topic-modelling}, last accessed 14.08.2026}

\begin{table}[h!]
    \small
    \centering
    \begin{tabular}{lcccc}
    \toprule
         \textbf{Configuration} & \textbf{$k$} & \textbf{$C_V$ (mean $\pm$ std)} & \textbf{TD (mean $\pm$ std)} & \textbf{Jaccard} \\
         \midrule
         \multirow{5}{*}{FASTopic Baseline}
             & 16 & \textbf{0.420 $\pm$ 0.058} & 1.000 $\pm$ 0.000 & 34.3\% \\
             & 20 & 0.399 $\pm$ 0.019 & 1.000 $\pm$ 0.000 & 29.3\% \\
             & 30 & 0.393 $\pm$ 0.015 & 0.999 $\pm$ 0.001 & 23.8\% \\
             & 40 & 0.373 $\pm$ 0.013 & 0.997 $\pm$ 0.004 & 25.3\% \\
             & 50 & 0.383 $\pm$ 0.011 & 0.990 $\pm$ 0.002 & 23.9\% \\
         \midrule
         \multirow{5}{*}{Seeded FASTopic}
             & 16 & \textbf{0.420 $\pm$ 0.018} & 1.000 $\pm$ 0.000 & 32.1\% \\
             & 20 & \textbf{0.419 $\pm$ 0.016} & 0.992 $\pm$ 0.011 & 39.0\% \\
             & 30 & 0.379 $\pm$ 0.015 & 0.937 $\pm$ 0.017 & 30.6\% \\
             & 40 & 0.381 $\pm$ 0.008 & 0.975 $\pm$ 0.010 & 26.7\% \\
             & 50 & 0.380 $\pm$ 0.014 & 0.982 $\pm$ 0.011 & 25.9\% \\
         \midrule
         \multirow{5}{*}{Regest FASTopic}
             & 16 & 0.412 $\pm$ 0.032 & 1.000 $\pm$ 0.000 & 33.4\% \\
             & 20 & 0.401 $\pm$ 0.023 & 1.000 $\pm$ 0.000 & 27.2\% \\
             & 30 & \textbf{0.401 $\pm$ 0.019} & 1.000 $\pm$ 0.000 & 26.7\% \\
             & 40 & 0.381 $\pm$ 0.027 & 0.999 $\pm$ 0.003 & 23.0\% \\
             & 50 & 0.366 $\pm$ 0.017 & 0.994 $\pm$ 0.006 & 24.5\% \\
         \midrule
         \multirow{5}{*}{Seeded+Regest FASTopic}
             & 16 & 0.411 $\pm$ 0.012 & 1.000 $\pm$ 0.000 & 42.7\% \\
             & 20 & 0.404 $\pm$ 0.027 & 0.995 $\pm$ 0.006 & 37.8\% \\
             & 30 & 0.391 $\pm$ 0.017 & 0.961 $\pm$ 0.017 & 25.9\% \\
             & 40 & \textbf{0.413 $\pm$ 0.021} & 0.996 $\pm$ 0.005 & 27.1\% \\
             & 50 & \textbf{0.389 $\pm$ 0.017} & 0.993 $\pm$ 0.006 & 24.3\% \\
         \midrule
        BERTopic & 3--22 & 0.475 $\pm$ 0.087 & 0.623 $\pm$ 0.136 & 39.1\% \\
         \bottomrule
    \end{tabular}
    \caption[FASTopic and BERTopic Evaluation]{Mean and standard deviation of $C_V$ coherence and topic diversity (TD) for FASTopic and BERTopic across five runs with different random initialisations. BERTopic's $k$ is data-driven and varies per run. Jaccard stability reports the mean best-match Jaccard similarity across all five runs in each configuration; higher values indicate more consistent topic recovery across initialisations. Best values among FASTopic configurations for $C_V$ at matched $k$ are in bold.}
    \label{tab:stability}
\end{table}

\begin{table}
\small
\centering
\begin{tabular}{ccc}
\toprule
\textbf{K} & \textbf{Mean Dominant Weight Range} \\
\midrule
16 & 0.0772 -- 0.0805 \\ 
\midrule
20 & 0.0620 -- 0.0658 \\ 
\midrule
30 & 0.0425 -- 0.0441 \\ 
\midrule
40 & 0.0331 -- 0.0345 \\ 
\midrule
50 & 0.0268 -- 0.0274 \\ 
\bottomrule
\end{tabular}
\caption[Topic Weights across FASTopic Runs]{Topic weights across all FASTopic runs (baseline, seeded, regest-augmented and seeded+regest-augmented)}
\label{tab:topic-distribution}
\end{table} 

\section{Discussion}
\subsection{Dense Retrieval} Overall, the dense retrieval results fall below the results reported in other intertextuality detection studies \parencite{caffagni2025benchmarking, schelb_loci_2026}. 
The performance gap is potentially attributable to several factors. First, the respective fine-tuning of the models for their target domain and task; second, the temporal linguistic distance between early modern Latin used by the Reformers and late/vulgar Latin of the Church Fathers; and third, the genre shift of epistolary exchange and patristic source texts. Cross-encoder reranking over the strongest bi-encoder top-100 improves recall for implicit references, consistent with prior findings that reranking improves retrieval quality \parencite{nogueira_multi-stage_2019, schelb_loci_2026}. However, cross-encoder inference is computationally expensive at full corpus scale. For explicit citations, dense retrieval falls ≈10-12\% points behind BM25, which also recovers implicit paraphrases. Dense retrieval contributes where vocabulary diverges as it finds implicit references that are semantically close to their source. Fusing both rankings via RRF therefore offers a competitive and less costly alternative to cross-encoder reranking, although on our evaluation data the gain amounts to only a few references.

\subsection{Topic Modelling}
The extent to which domain resources from \textit{Bullinger Digital} yield gains in topic coherence is difficult to assess. Seed guidance by the categories and regest augmentation both inform the topic structure, while certain limitations persist across all configurations.

\begin{figure}
    \centering
    \includegraphics[width=1.0\textwidth]{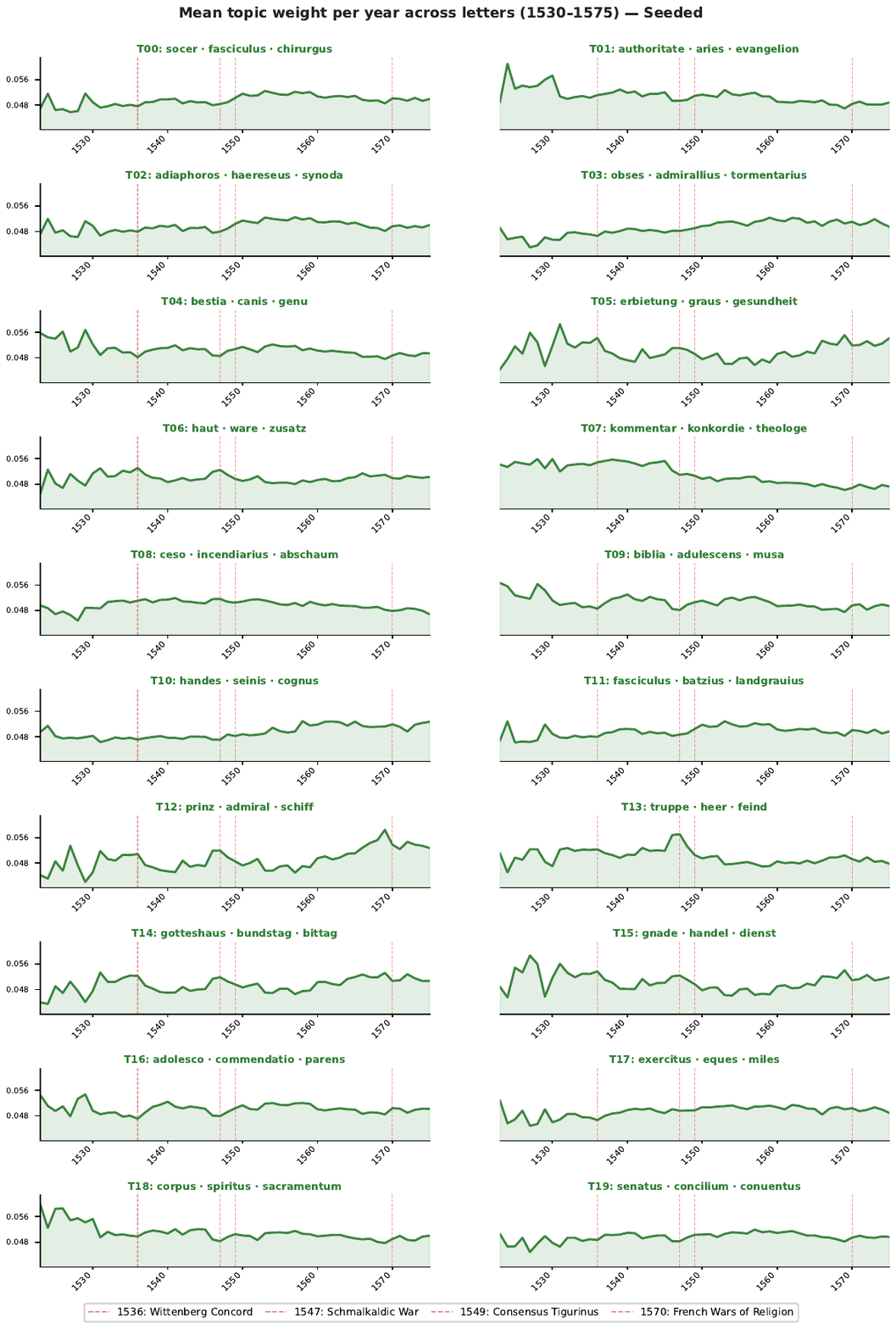}
    \caption[Topics over Time, Seeded Configuration ($K=20$)]{Mean topic weight per year across BDC letters for the seeded configuration ($K=20$) over the correspondence period 1530--1575. Vertical dashed lines mark four historically significant events: the Wittenberg Concord (1536), the Schmalkaldic War (1547), the \textit{Consensus Tigurinus} (1549), and the French Wars of Religion (dated here to 1570).}
    \label{fig:seed-time-appendix}
\end{figure}

Several topics discovered with the enhanced topic models exhibit a degree of theological specificity that exceeds broad thematic categorisation and reflects identifiable sub-discourses within the correspondence. One example is the anti-papal apocalyptic topic found by Seeded FASTopic at $k=20$ (cf. Figure \ref{fig:seed-time-appendix}) (T04: \textit{bestia, canis, antichristus, diabolus, papatus, persecutio, grex, ouis, monstrum, tentatio}). These terms map onto how Reformists thought of the papacy as the Antichrist of Revelation \parencite[p. 217-218]{Almond_2020}, with \textit{grex} and \textit{ouis} (flock and sheep) invoking the image of the popular Shepherd image of John~10 used to frame the persecuted Reformists \parencite[p. 82-83]{BullDa1966}. Another one is T08 at $k=20$, a confessional conflict topic (\textit{ceso, incendiarius, abschaum, kanoniker, buceraner, indictio, catabaptistus, religionsfriede, vorfall, vorzeichen}) that captures how the letters frame disturbances by classifying actors such as Anabaptists (\textit{catabaptistus}), Catholic clergy (\textit{kanoniker}), and Bucerian reformers (\textit{buceraner}) -- opponents often described with polemics in Bullinger's network \parencite[p. 12]{campi_heinrich_2004} -- within moralised categories (\textit{abschaum, incendiarius}), while embedding these disputes in the context of confessional conflicts (\textit{religionsfriede}).

\begin{figure}
    \centering
    \includegraphics[width=1.0\textwidth]{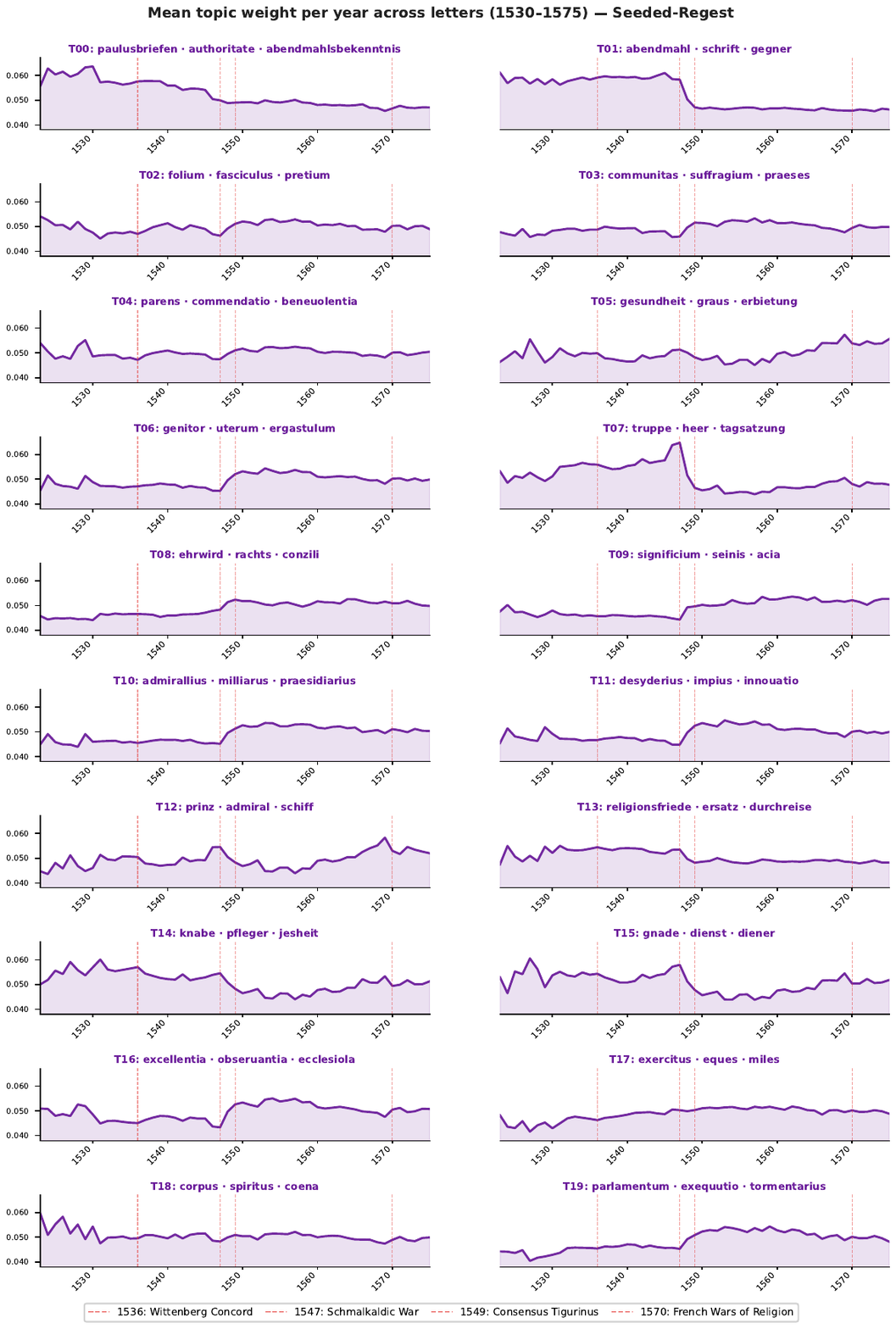}
    \caption[Topics over Time, Seeded+Regest Configuration ($K=20$)]{Mean topic weight per year across BDC letters for the Seeded+Regest configuration ($K=20$) over the correspondence period 1530--1575. Vertical dashed lines mark four historically significant events: the Wittenberg Concord (1536), the Schmalkaldic War (1547), the \textit{Consensus Tigurinus} (1549), and the French Wars of Religion (dated here to 1570).}
    \label{fig:regest-time}
\end{figure}

Temporal topic distributions offer partial validation of thematic coherence. The clearest case is the military topics. A troop topic peaks at the Schmalkaldic War (1546--1547) both with regests (T07, Figure~\ref{fig:regest-time}) and without them (T13, Figure~\ref{fig:seed-time-appendix}). A naval topic (T12) rises in both configurations towards the French Wars of Religion and the Dutch Revolt (around 1560--1570). In the Seeded+Regest configuration, the Eucharist topics (T00, T01) decline around 1549, which would fit the \textit{Consensus Tigurinus}. However this also coincides with the end of regest coverage in the 1540s. These shifts could therefore also reflect the change in document representation rather than historical developments. The same applies to the near-identical plots of topics with unrelated vocabulary (e.g., T06, T08, T11, T16), which we read as noise. Distributions that recur across configurations with and without regests thus offer a more reliable criterion for coherent, evolving topics than coherence and diversity metrics alone. 

The absence of several major Reformation topics should also be noted. An experimental Latin-only run, encoding only the Latin portion of each letter and constructing the BoW vocabulary exclusively from Latin noun lemmas, revealed finer-grained doctrinal topic structures at $k = 20$ than any of the four multilingual configurations reported. Some examples are a sacrament-discussion cluster (\textit{glorificatio, signaculum, expiatio, sacrifitium, sumptio,~...}), a Trinity topic (\textit{trinitas, mediator, doctrina, persona, deus, consensio, substantia, coelum, natura,~...}), an eschatological topic (\textit{ecclesiola, coelestis, immortalitas, letitia, ueneratio,~...}), and an ecclesiological topic (\textit{excommunicatio, scandalum, aedificatio, uocatio},~...). This suggests that the ENHG administrative, political, and everyday vocabulary within the joint topic space suppresses topics of theological discourse.

An experimental concept modelling run using LLooM on 50 random letters shows a more interpretable approach in comparison to the FASTopic outputs.\footnote{\url{https://github.com/len-rtz/bullinger-concept-modelling}, last accessed 14.08.2026} Applied to the pilot sample, a distillation step condensed the letters into 177 bullet-point summaries, which were clustered (HDBSCAN) into eight thematic groups. From these clusters, a synthesis step proposed 59 candidate concepts. Each is generated as an inclusion criterion, unlike other approaches that produce a keyword list. For example: \enquote{Does the text argue against the authority of the Pope or the Roman papacy in favour of national or godly law?} for \textit{Anti-Papal Authority}. Output of a comparable topic in a FASTopic run looks like this: \textit{bestia, ouis, antichristus, diabolus, papatus, persecutio, grex, exilium, monstrum, tentatio}. An automatic review pass in LLooM then merged eight overlapping concept pairs. This internal consistency-checking could also be skipped and done manually to proceed with concept scoring. We selected ten theological concepts for concept scoring across the pilot sample (e.g., \textit{Justification by Faith}, \textit{Scriptural Supremacy}, \textit{Sacramental Integrity}, \textit{Anti-Lutheran Polemics}). For each letter–concept pair, LLooM's output is a scoring criterion (a generated prompt), a score, and a rationale justifying the score, making it considerably more interpretable than, e.g., FASTopic. We illustrate this with the example of the letter HBBW 12, Nr.~1597 (p.~30)\nocite{hbbw12}, in which Kaspar Schwenckfeld, a German Reformer, is sending the Zurich reformers a copy of his treatise on the glory of Christ, urging them to weigh it against the Scripture and arguing that full doctrinal understanding is revealed incrementally rather than all at once. LLooM gives the concept \textit{Sacramental Integrity} (scoring prompt: \enquote{Does the text address the validity or meaning of the sacraments?}) a low score of 0.25, with the rationale correctly noting that \enquote{The text discusses the nature of Christ's glory, the sufficiency of scripture, and the process of spiritual revelation, but it does not specifically address the validity or theological meaning of the sacraments.} -- a plausible rejection. By contrast, \textit{Spiritual Presence} (scoring prompt: \enquote{Does the text argue for the presence of Christ through faith and the spirit rather than physical location?}) is scored 1.00, with the rationale identifying that the letter \enquote{argues that Christ's glory and knowledge are revealed through the Holy Spirit and faith rather than through human craft or physical presence}. Here, the generated rationale diverges from the letter's content because in it, Schwenckfeld never addresses where or how Christ is present relative to physical location. The level of interpretability of the results comes at a computational cost. Inference ran locally on a MacBook (M1) via Turbofieldflare using \nolinkurl{gemma-4-26b-a4b-it}. Concept modelling the 50-letter sample took took 15 hours, a cost scaling linearly with corpus size. We therefore report the pilot as exploratory groundwork for future work.

\section{Limitations}
Regarding text reuse, a first limitation concerns the interpretability of the encoder representations underlying the retrieval pipeline. As \textcite{taddei_detecting_2025} note, computational models \enquote{cannot distinguish between coincidental similarity, shared literary tradition, and intentional allusion} (p. 8); passages surfaced by the system require humanistic interpretation before any grounded claim of patristic influence can be sustained. Following \textcite{riemenschneider2023graecia}, the pipeline is best understood as a first-pass exploration tool. Ground truth evaluation is, further to the limitations discussed in Section \ref{sec:evaluationdata}, constrained by the size of the evaluation dataset, with $n=100$ limiting the statistical robustness of the reported retrieval metrics. The absence of a false positive evaluation, an assessment of how often the pipeline retrieves passages that are superficially similar but philologically unrelated, means that precision metrics cannot be estimated. It is also worth noting that multilingual references remain an open question, as only one out of nine non-Latin-to-Latin evaluation reference was recovered by the multilingual bi-encoder.

Several limitations also bear on the interpretation of the topic modelling results. Persistent junk clusters appear across all four configurations with overlapping vocabulary. The degree to which seeding and regest augmentation construct or alter topic structure cannot be cleanly disentangled. The regest interpolation weight of 0.5 was set for practical reasons without systematic tuning, and the direction of the bias it introduces, whether it over- or under-weights scholarly context, is unknown. 

\section{Future Work}
Several directions present themselves for future work. Fine-tuning the language models used for retrieval on domain-specific annotations following \textcite{caffagni2025benchmarking, schelb_loci_2026} and others represents a natural next step for enhancing dense retrieval. Training data for that could be readily constructed from the numerous references annotated in the HBBW. Further, the role of named-entity information from the letters and further integration of feature/metadata deserve systematic investigation \parencite{Romanello_Citation_2016} as direct patristic references are frequently mediated through person mentions. Segmentation strategy equally warrants empirical scrutiny, as does the introduction of metadata-enriched segmentation \parencite{lau_mining_2024}, such as citation markers or discourse cues, which may improve retrieval performance over the current segmentation. Though costly, LLM-assisted reference detection or interpretation \parencite{umphrey_investigating_2024, fu_ancienttrd_2025} offers an alternative or complementary approach, provided hallucination risks are managed. 

As already brought up in the discussion, topic modelling could be explored further along several axes. One option would be to incorporate the per-letter \textit{Bullinger Digital} category tags as a signal instead of using them only as topic seeds. The effect of restricting the vocabulary to Latin only versus ENHG only could likewise be investigated further. A direction worth exploring is also a finer-grained application of LLooM to the corpus. Our pilot applied concept induction at the letter level, where documents often exceed several hundred words and address multiple topics at once (reflected in the 177 bullet points generated from only 50 letters). \textcite{lam2024conceptinduction}'s original validation was done on short, self-contained, topically coherent social media posts. Feeding full letters into the pipeline may therefore understate LLooM's capacity for specific, idea-level concepts. Since a subset of letters includes scholarly regest annotations in which regest paragraphs align with specific paragraph-level spans in the letter body (marked via \texttt{div/@corresp} in the TEI encoding), segmenting along these boundaries would yield shorter, more topically coherent input closer to LLooM's intended granularity. Following the same logic behind using regest embeddings in FASTopic, future work should compare concept induction on regest-segmented text against the same segments augmented with their corresponding regest text, to isolate whether segmentation, regest augmentation, or their combination improves concept specificity and interpretability.

Beyond detection at the level of individual letters, a natural extension is tracing whether patristic references and the ideas they carry propagate through Bullinger's correspondence network. For example, one could examine whether a shared citation coincides with specific topics/correspondents or reading recommendations. Such network-level diffusion analysis would require linking detected references to sender–recipient metadata and modelling their spread over time \parencite{roller_tracing_2023}.

More generally, a cross-corpus evaluation on the ThBW could assess the methodological generalisability beyond the Bullinger corpus. Most importantly, future work should involve theologians and religious scholars who can provide qualitative feedback.

\section{Conclusion}
We investigated the degree to which computational methods can detect patristic references and model topics as proxies to theological ideas in Heinrich Bullinger's 16th-century correspondence. The best-performing domain-adapted Latin encoder, \nolinkurl{LaBERTa-W_VULG-S_VL}, performs better than other bi-encoders, while BM25 beats all dense approaches on explicit citations. The hybrid RRF improves implicit recall over either method alone at minimal additional cost.
Topic modelling over the \textit{Bullinger Digital} corpus with FASTopic recovers thematically distinct and, in some cases, historically coherent patterns. Neither metadata-informed configuration is reliably superior to unmodified FASTopic. Gains remain difficult to assess. Although we aimed to use topic-letter distributions to potentially inform reference retrieval, dominant topic weights barely exceed a uniform distribution, suggesting that the current topic representations are too diffuse to inform retrieval. Nevertheless, connecting retrieval with topics remains the intended roadmap for future work once tighter topic modelling approaches are established. A pilot application of LLooM suggests a more interpretable, criterion-based alternative to keyword topics at additional computational cost. 
Together, these results position patristic reference detection as tractable for explicit citation but harder for implicit references, with hybrid dense and sparse retrieval and future domain fine-tuning as promising paths to narrow that gap. Topic modelling offers a productive but still coarse-grained proxy for the ideas discussed during the Reformation. Both point to the potential of computational methods to facilitate large-scale exploration of Bullinger’s correspondence, while highlighting the need for further methodological refinement, more robust evaluation and philological validation to establish more reliable connections between patristic references and theological ideas.

\newpage

\printbibliography

\section*{AI Statement}
During the preparation of this work, the authors used Claude Sonnet 4.5 in VS-Code in order to assist in coding. The authors reviewed and edited the scripts as needed and take full responsibility for the content of the published article.
\newpage

\appendix

\section{Appendix Section} \label{appdx:first}
\subsection{Initialised Seed Topics}
\label{append:seedtopics}
During the project \textit{Bullinger Digital}, the correspondence was thematically annotated combining expert-defined topic sets, manual annotation of a small sample, and LLM prompting with the eleven categories\footnote{\url{https://github.com/bullinger-digital/bullinger-auxiliary-scripts/blob/main/Verschlagwortung/topics/topics4_en.json}, last accessed 25.05.2026} displayed in Table \ref{BD-categories}. The seed phrases used to initialise each topic embedding were derived from these. Load-bearing phrases were extracted manually from each description and supplemented with thematically aligned terminology. Some categories were split into multiple topic slots to account for their greater internal diversity and importance in the context of theological discourse, as detailed in Table~\ref{tab:seed_phrases}.

\begin{table}
\small
\centering
\begin{tabular}{|p{5cm}|p{10cm}|}
\hline
\textbf{Category} & \textbf{Description} \\
\hline
Faith and Doctrine in Theory & Theological and doctrinal discussions defining Reformed beliefs, \textbf{biblical interpretation}, \textbf{predestination}, \textbf{church discipline}, and disputes such as the \textbf{Eucharist controversy}. Focuses on belief systems rather than practice. \\
\hline
Faith and Doctrine in Practice & Application of religious doctrine in daily life and \textbf{church governance}, including \textbf{worship practices}, \textbf{rituals}, \textbf{clergy roles}, \textbf{sacraments}, and \textbf{ceremonies}. Focuses on institutional and behavioral consequences of doctrine. \\
\hline
Military and Political Affairs & Reports and analyses of \textbf{wars}, \textbf{military actions}, \textbf{diplomatic alliances}, \textbf{political strategies}, and \textbf{governmental decisions}. \\
\hline
Religious Persecution and Martyrdom & Discussions of Protestant persecution, including \textbf{exile}, \textbf{imprisonment}, \textbf{execution}, \textbf{martyrdom}, \textbf{massacres}, and fears or experiences of persecution. \\
\hline
Education System & Covers discussions on \textbf{educational institutions}, \textbf{recruitment and placement of scholars}, \textbf{theological} and \textbf{humanist teaching}, \textbf{curriculum} of \textbf{schools} and academies, funding and establishment of schools, exchanges between scholars, and \textbf{university} organization. \\
\hline
Books, Printing, and Publishing & Discussions about \textbf{printing books}, interactions with \textbf{printers}, availability of \textbf{printed volumes}, and the \textbf{circulation} and influence of \textbf{printed works}. \\
\hline
Requests and Petitions & Explicit or implicit requests for action, \textbf{support}, \textbf{guidance}, assistance, or \textbf{favors} directed at specific individuals or groups. This includes \textbf{appeals for help}, efforts to exert influence, or attempts to activate \textbf{personal or political networks}. Excludes prayers, hopes for divine intervention, and general communication. \\
\hline
Everyday Life & Daily life events, routines, and mundane activities, including food, \textbf{cooking}, \textbf{buying and selling}, going to \textbf{the market}, doing things around \textbf{the house}, etc. \\
\hline
Disasters and Natural Phenomena & Extraordinary natural events, disasters, and phenomena such as \textbf{harvest failures}, \textbf{meteorological observations}, \textbf{celestial events}, \textbf{fires}, and \textbf{floods}. \\
\hline
Illness, Death, and Recovery in Society & Societal descriptions of \textbf{plague} or other illnesses, \textbf{recovery}, and \textbf{nonviolent death} due to illness, old age, or childbirth. Excludes violent or war-related deaths. \\
\hline
Illness, Death, and Recovery among Family and Close Friends & Personal accounts of \textbf{suffering}, \textbf{mourning}, \textbf{illness}, or \textbf{death} within the \textbf{household} or among \textbf{close friends}, including \textbf{recovery} and healing experiences. \\
\hline
\end{tabular}
\caption[The Eleven Bullinger Digital Categories]{The eleven categories in Bullinger Digital. Text in bold was the base for the category topic seed(s).}
\label{BD-categories}
\end{table}

\begin{table}
\small
\centering
\begin{tabular}{|c|p{3cm}|p{10cm}|}
\hline
\textbf{Topic} & \textbf{Category} & \textbf{Seed Terms/Phrases} \\
\hline
0 & \multirow{3}{3cm}{Faith and Doctrine in Theory} & \textbf{predestination}, \textbf{divine election}, double predestination, \textbf{God's decree}, reprobation, providence, foreknowledge, eternal counsel \\
\cline{1-1}\cline{3-3}
1 &  & \textbf{Christ's presence in the Eucharist}, \textbf{Eucharist}, \textbf{Abendmahlstreit}, real presence, memorialism, consubstantiation, Lord's Supper, sacramental union, spiritual manducation \\
\cline{1-1}\cline{3-3}
2 &  & \textbf{sola scriptura}, \textbf{biblical interpretation}, scriptural authority, \textbf{church governance}, \textbf{church discipline}, conciliarism, canon, hermeneutics, exegesis \\
\hline
3 & \multirow{2}{3cm}{Faith and Doctrine in Practice} & \textbf{worship practices}, \textbf{clerical vestments}, \textbf{rituals}, \textbf{church ceremonies}, liturgical reform, rites, church order \\
\cline{1-1}\cline{3-3}
4 &  & \textbf{ecclesiastical office}, \textbf{clergy roles}, \textbf{church discipline}, \textbf{pastor appointment}, \textbf{excommunication}, visitation, ordination, synod \\
\hline
5 & \multirow{2}{3cm}{Military and Political Affairs} & \textbf{diplomatic alliance}, \textbf{military campaign}, \textbf{war}, soldier, siege, mercenary, truce \\
\cline{1-1}\cline{3-3}
6 &  & \textbf{ruler's policy}, \textbf{governmental decision}, \textbf{political strategy}, \textbf{imperial diet}, treaty, council, magistrate \\
\hline
7 & \multirow{2}{3cm}{Religious Persecution and Martyrdom} & \textbf{Protestant exile}, \textbf{expulsion}, flight from persecution, \textbf{refugee}, displacement, banishment \\
\cline{1-1}\cline{3-3}
8 & & \textbf{martyrdom}, \textbf{execution}, \textbf{imprisonment}, \textbf{massacre}, Inquisition, suffering for faith \\
\hline
9 & Education System & \textbf{education}, \textbf{school foundation}, \textbf{university}, \textbf{scholar recruitment}, \textbf{theological teaching}, \textbf{humanist teaching}, curriculum, stipend, gymnasium, lectorship \\
\hline
10 & Books, Printing, Publishing & \textbf{printing books}, \textbf{book circulation}, \textbf{printing press}, \textbf{printer}, published works, manuscript, edition \\
\hline
11 & Requests and Petitions & \textbf{appeal for support}, appeal for guidance, \textbf{personal favour}, \textbf{political network}, \textbf{petition}, intercession, recommendation, patronage \\
\hline
12 & Everyday Life & \textbf{everyday life}, \textbf{cooking}, \textbf{market}, \textbf{food trade}, \textbf{household}, \textbf{buying and selling}, cleaning, provisions \\
\hline
13 & Disasters and Natural Phenomena & \textbf{harvest failure}, \textbf{flood}, \textbf{celestial event}, \textbf{fire}, \textbf{meteorological observation}, natural phenomena, disaster, comet, earthquake, famine \\
\hline
14 & Illness, Death in Society & \textbf{plague}, epidemic, \textbf{societal mortality}, \textbf{recovery}, pestilence, mortality, contagion \\
\hline
15 & Illness, Death among Family & \textbf{personal grief}, \textbf{household sickness}, \textbf{death of loved one}, \textbf{mourning}, bereavement, loss, consolation \\
\hline
\end{tabular}
\caption[Seed Terms and Phrases for Centroid Construction]{Seed terms and phrases used to find neighbours in the embedding space for initialising a topic centroid in Seeded-FASTopic, grouped by topic slot. Bold terms/phrases were derived directly from the Bullinger Digital category descriptions; remaining ones were added.}
\label{tab:seed_phrases}
\end{table}

\end{document}